\documentclass[twocolumn,journal]{IEEEtran}
\usepackage{graphicx}
\usepackage{balance}
\usepackage{amsmath}
\usepackage{amssymb}
\usepackage{color}
\usepackage{multicol}
\usepackage{amsthm}
\usepackage[usenames,dvipsnames]{xcolor}

\usepackage{cite}
\usepackage{float}

    \newcommand\blfootnote[1]{%
  \begingroup
  \renewcommand\thefootnote{}\footnote{#1}%
  \addtocounter{footnote}{-1}%
  \endgroup
}

\begin{document}
\title{Probabilistic Band-Splitting for a Buffered Cooperative Cognitive Terminal}
\author{ Ahmed El Shafie$^\dagger$, Ahmed Sultan$^\star$, Tamer Khattab$^*$\\
\small \begin{tabular}{c}
$^\dagger$Wireless Intelligent Networks Center (WINC), Nile University, Giza, Egypt. \\
$^\star$Department of Electrical Engineering, Alexandria University, Alexandria, Egypt.\\
$^*$Electrical Engineering, Qatar University, Doha, Qatar. \\
\end{tabular}
}

\date{}
\maketitle

\blfootnote{This paper was made possible by a NPRP grant 6-1326-2-532 from the
Qatar National Research Fund (a member of The Qatar Foundation). The
statements made herein are solely the responsibility of the authors.}
\thispagestyle{empty}
\pagestyle{empty}
\begin{abstract}
In this paper, we propose a cognitive protocol that involves cooperation between the primary and secondary users. In addition to its own queue, the secondary user (SU) has a queue to store, and then relay, the undelivered primary packets. When the primary queue is nonempty, the SU remains idle and attempts to decode the primary packet. When the primary queue is empty, the SU splits the total channel bandwidth into two orthogonal subbands and assigns each to a queue probabilistically. We show the advantage of the proposed protocol over the prioritized cognitive relaying (PCR) protocol in which the SU assigns a priority in transmission to the primary packets over its own packets. We present two problem formulations, one based on throughput and the other on delay. Both optimization problems are shown to be linear programs for a given bandwidth assignment. Numerical results demonstrate the benefits of the proposed protocol.
\end{abstract}
\section{Introduction}
\IEEEPARstart{C}\small{ooperation} between different nodes in a wireless communication network can efficiently increase the achievable transmission rate of each node. In the context of cognitive radios, cooperation has been investigated in many papers such as \cite{simeone,khattab,erph,su2011active}. Simeone {\it et al.} \cite{simeone} investigated the maximum stable throughput of a secondary transmitter that relays the undelivered primary packets. The secondary user (SU) adapts its transmit power to maximize its stable throughput.
In \cite{khattab}, the secondary transmitter relays a certain fraction of the primary undelivered packets to minimize the average secondary queueing delay subject to a power
budget for the relayed primary packets. The authors considered a prioritized cognitive relaying (PCR) protocol in which transmission priority is assigned to the relaying queue over the secondary own data queue. Specifically, the SU cannot transmit any of its own packets until both the primary queue and the relaying queue become empty. Kompella {\it et al.} \cite{erph} characterized the stable-throughput region of a network composed of one primary user (PU) and one cooperating SU with multipacket reception (MPR) capability at the receiving terminals. The authors of \cite{su2011active} proposed a new cooperative protocol for bufferless terminals. Every time slot, part of the PU's time slot duration and bandwidth are being released to the SU. Specifically, portion of the primary bandwidth is released for the cognitive radio (CR) user and {\it half} of its time slot duration. During the first half of the time slot, the SU receives the PU data. Then, it amplifies-and-forwards the received packet during the remainder of the time slot. At the primary destination, the signal from the PU
transmitted over the first half of the time slot and the forwarded signal from the SU during the second half of the time slot
are jointly decoded using maximum-ratio combining (MRC). The users are assumed to have a complete knowledge of the instantaneous transmit channel state information (CSI).

In \cite{rong2012cooperative}, the authors studied the impact of cooperation in a wireless multiple-access system. A packet from any of the sources is delivered to the common destination through either
a direct link or through cooperative relaying by intermediate
source nodes. The authors investigated the PCR protocol in which a node with lower priority of transmission must deliver the relaying packets of the higher priority nodes before transmitting its own packets. The stability region and queueing delays were characterized.

 In this paper, we consider a cooperative cognitive scenario with one primary transmitter-receiver pair and one secondary transmitter-receiver pair. In addition to its own queue, the SU maintains a queue to store, and then relay, the undelivered primary packets. When the primary queue is nonempty, the SU remains idle and attempts to decode the primary packet. When the primary queue is empty, the SU splits the total channel bandwidth into two, generally unequal, portions and assigns each to a queue probabilistically at each time slot.

The contributions of this paper can be summarized as follows. To the best of our knowledge, the analysis of bandwidth splitting for a buffered SU, which sends its own data and helps relay some of the data of a buffered PU, is reported in this paper for the first time. Moreover, we propose a probabilistic assignment of subbands to the secondary and relaying queues. The proposed protocol is simple and does not require the knowledge of the CSI at the transmitters. In addition to the problem formulation based on throughput, as in several other works, we also investigate the queueing delays and provide a formulation based on minimizing the secondary delay under the constraint that the primary delay does not exceed a specified threshold. We show that the proposed protocol outperforms the PCR protocol.

This paper is organized as follows. In the following Section, we discuss the system model adopted in this paper. In Section \ref{sec2}, we describe the proposed protocol and present the problems formulations. We present some numerical results of the optimization problems presented in this paper in Section \ref{sec3}. In Section \ref{sec4}, we conclude the paper.
\section{SYSTEM MODEL} \label{w100}
\begin{figure}
  \centering
  \includegraphics[width=1\columnwidth]{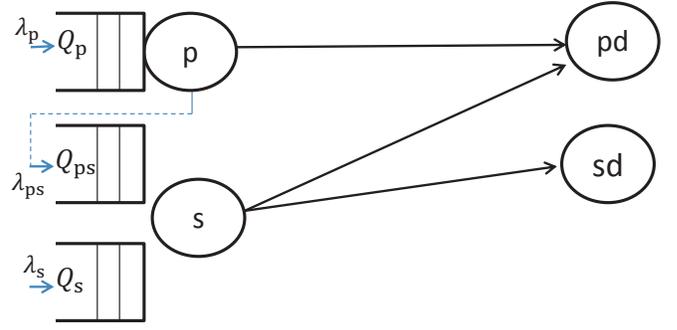}\\
   \caption{Primary and secondary links and queues. The solid links represent the communication links between nodes.
}\label{fig0}
\end{figure}

 We assume a simple configuration consisting of one primary transmitter `${\rm p}$', one secondary transmitter `${\rm s}$', one primary destination $`{\rm pd}$' and one secondary destination $`{\rm sd}$' (as shown in Fig. \ref{fig0}). This can be seen as a part of a lager network with multiple primary bands operating under frequency division multiple-access (FDMA). Each band is composed of one primary transmitter-receiver pair and one secondary transmitter-receiver pair. Time is slotted and each slot is of $T$ seconds in length. The bandwidth assigned to each primary transmitter is $W$ Hz. For simplicity in presentation, we provide the analysis of one of those orthogonal frequency bands.

Each terminal has an infinite buffer for storing its own arrived packets, denoted by $Q_\ell$, $\ell\in\{{\rm p,s}\}$, `${\rm p}$' for primary and `${\rm s}$' for secondary. The SU has an additional infinite capacity buffer for storing the primary relaying packets, denoted by $Q_{\rm ps}$. Each data packet contains $b$ bits. The arrivals at $Q_{\ell}$ are assumed to be independent and identically distributed Bernoulli random variables from slot to slot with mean arrival rate $\lambda_{\ell}\!\in\![0,1]$ packets per time slot. Arrivals are also independent from terminal to terminal.

Each destination sends a feedback message to inform the respective transmitter about the decodability status of the transmitted packet. The retransmission mechanism is based on the feedback acknowledgement/negative-acknowledgement (ACK/NACK) messages. If a packet is decoded properly at the respective destination, an ACK is fed back to the respective transmitter. On the other hand, if the destination cannot decode the transmitted packet, a NACK message is fed back to the respective transmitter. If the primary destination cannot decode the primary packet but the SU can, the SU feeds back the primary transmitter with an ACK, and the packet is dropped from the PU's queue. Due to the broadcast nature of the wireless channel, all nodes in the system can hear the feedback ACK/NACK messages. Hence, the SU overhears the primary feedback signal. The overhead for transmitting the ACK and NACK messages is assumed to be very small
compared to data packet sizes. Furthermore, errors in packet feedback acknowledgement are negligible due to the use of
strong channel codes \cite{sadek}.

The channel gains are assumed to remain constant over the duration of the time slot and the band of operation. The channel
is assumed to be known perfectly only at the receivers. Let $h^t_{j,k}$ (for the $j\!\rightarrow\!k$ link) denote the channel gain between transmitting node $j$ and receiving node $k$ at time slot $t$, where $j \in \{{\rm s},{\rm p}\}$, $k \in \{{\rm s},{\rm sd},{\rm pd}\}$ and $j\ne k$, which is exponentially distributed in case of Rayleigh fading channel with mean $\sigma_{j,k}$. Hereinafter, the time notation is omitted from all symbols for simplicity. Channel gains are independent from slot to slot and link to link. Thermal noise at receiving nodes is modeled as a complex additive white Gaussian noise (AWGN) with zero mean and power $\mathcal{N}_\circ$ Watts per unit frequency (Watts/Hz). Transmitter $j$ is assumed to transmit with power $P^{j}$ Watts/Hz .
\noindent Outage occurs when the transmission rate exceeds the channel capacity. The probability of channel outage of the link between node $j$ and node $k$ is given by \cite{simeone}
\begin{equation}
{\rm Pr}\bigg\{O_{j,k}\bigg\}=\mathbb{P}_{j,k}={\rm Pr}\biggr\{r_j > W_{j,k} \log_{2}\left(1+\alpha_{j,k}\right)\biggr\}
\end{equation}
\noindent where $O_{j,k}$ is the event of channel outage between node $j$ and node $k$, ${\rm Pr}\big\{O_{j,k}\big\}$ is the probability of the argument event $O_{j,k}$, $r_j$ is the transmission rate of transmitter $j$, $W_{j,k}$ is the transmission bandwidth used for the communication between node $j$ and node $k$, $\alpha_{j,k}=h_{j,k}\gamma_{j,k}$ is the received signal-to-noise-ratio (SNR) at node $k$, and $\gamma_{j,k}\!=\!P^{j}/\mathcal{N}_\circ$. The outage probability can be rewritten as
\begin{equation}
\mathbb{P}_{j,k}\!=\!{\rm Pr}\Biggr\{h_{j,k}<\frac{2^{\frac{r_j}{W_{j,k}}}\!-\!1}{\gamma_{j,k}}\Biggr\}\!=\!1\!-\!\exp\bigg(-\frac{2^{\frac{r_j}{W_{j,k}}}-1}{\gamma_{j,k}\sigma_{j,k}}\bigg)
\label{ghj}
\end{equation}
The link $j\rightarrow k$ is not in outage with probability
\begin{equation}
\overline{\mathbb{P}_{j,k}}\!=\!1\!-\!\mathbb{P}_{j,k}\!=\!\exp\bigg(-\frac{2^{\frac{r_j}{W_{j,k}}}-1}{\gamma_{j,k}\sigma_{j,k}}\bigg)
\label{ghj}
\end{equation}

 The SU senses the channel for $\tau$ seconds to discern the state of the PU's queue. The sensing outcome is assumed to be perfect as in, e.g., \cite{khattab} and \cite{rong2012cooperative}. Since the number of bits in a packet is $b$, the transmission rate of the secondary transmitter is then given by
\begin{equation}
r_{\rm s}=\frac{b}{T-\tau}
\label{r_s}
\end{equation}
where $T-\tau$ in (\ref{r_s}) is the remaining time for data transmission after channel sensing. On the other hand, since the PU transmits its packet whose length is $b$ bits over the whole slot duration, $T$, the data transmission rate of the PU is given by
\begin{equation}
r_{\rm p}=\frac{b}{T}
\label{r_p}
\end{equation}

\section{Proposed Cooperative Cognitive Protocol}\label{sec2}
\begin{figure}
  \centering
  \includegraphics[width=1.05\columnwidth]{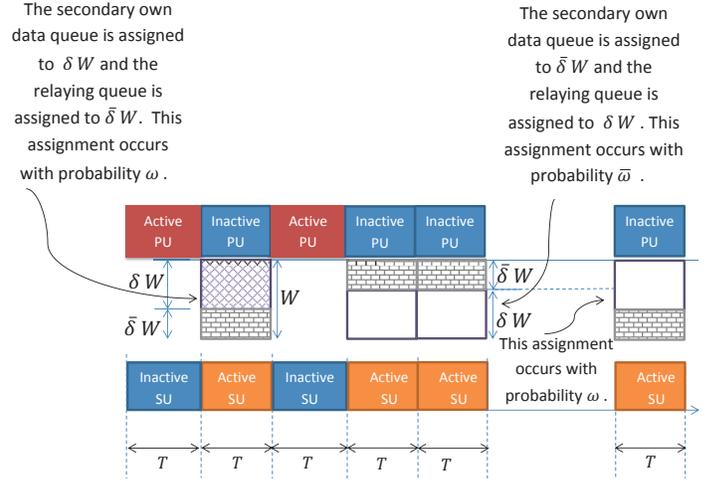}\\
   \caption{Time slot structure and system operation. In the figure, $\overline{\mathcal{X}}=1-\mathcal{X}$.
}\label{fig01}
\end{figure}

In this section, we analyze the proposed cooperative cognitive protocol. Under the cooperative protocol, the SU's operation can be summarized as follows. At the beginning of the time slot, the SU senses the channel for $\tau$ seconds to detect the state of primary activity. When the PU is active, the SU remains silent and attempts to decode the primary packet and store it if the primary destination fails to decode it. When the PU is inactive, the SU splits the total bandwidth of the channel into two orthogonal subbands, and sends a packet from each of its queues. Assume that the SU splits the overall bandwidth into two orthogonal subbands, $W_{\rm 1}\!=\!\delta_1 W\!=\!\delta W$ and $W_{\rm 2}\!=\!\delta_2W\!=\!(1\!-\!\delta) W$ with $\delta_1\!+\!\delta_2\!=\!1$ and $W_1\!+\!W_2\!=\!W$.\footnote{Equivalently, we can divide the time available for secondary transmission, $T\!-\!\tau$, into $T_1\!=\!\delta (T\!-\!\tau)$ and $T_2\!=\!(1\!-\!\delta) (T\!-\!\tau)$, where $T_1\!+\!T_2\!=\!T-\tau$.} At each sensed free time slot, the SU assigns $\delta W$ to $Q_{\rm s}$ and $(1\!-\!\delta) W$ to $Q_{\rm ps}$ with probability $\omega$; or assigns $\delta W$ to $Q_{\rm ps}$ and $(1\!-\!\delta) W$ to $Q_{\rm s}$ with probability $1\!-\!\omega$.\footnote{The proposed protocol is an inner bound to a protocol that assigns the whole bandwidth to a nonempty queue when the other queue is empty. The general protocol couples the queues and makes the analysis intractable.} The system operation is shown in Fig. \ref{fig01}.

A packet at the head of $Q_{\rm p}$ is served with probability one minus the probability that the links ${\rm p}\!\rightarrow\!{\rm pd}$ and ${\rm p}\!\rightarrow\!{\rm s}$ are being in outage simultaneously. That is,
\begin{equation}
\begin{split}
\label{xor}
 \mu_{\rm p} &= 1-\mathbb{P}_{{\rm p},{\rm s}} \mathbb{P}_{\rm p,pd}
   \end{split}
 \end{equation}
 The probability that the PU's queue being empty is given by \cite{rong2012cooperative,sadek}
 \begin{equation}
\begin{split}
 {\rm Pr}\{Q_{\rm p}=0\}\!=\!\pi_\circ &= 1-\frac{\lambda_{\rm p}}{\mu_{\rm p}}
   \end{split}
 \end{equation}

 A packet is arrived at $Q_{\rm ps}$ when the PU's queue is nonempty, the link ${\rm p}\!\rightarrow\!{\rm pd}$ is in outage, and the link ${\rm p}\!\rightarrow\!{\rm s}$ is not in outage. This can be written as
 \begin{equation}
\begin{split}
\label{lambdaps}
   \lambda_{\rm ps}&=  \mathbb{P}_{\rm p,pd}\overline{\mathbb{P}_{{\rm p},{\rm s}}} \overline{\pi_\circ}
     \end{split}
 \end{equation}
 where $\overline{\mathcal{X}}\!=\!1\!-\!\mathcal{X}$. When the PU is inactive, a packet at the head of $Q_{\rm s}$ is served in either one of the following events. If $Q_{\rm s}$ is assigned to $\delta W$, which occurs with probability $\omega$; and the link ${\rm s}\!\rightarrow\!{\rm sd}$ is not in outage. Or if $Q_{\rm s}$ is assigned to $(1\!-\!\delta) W$, which occurs with probability $1\!-\!\omega$; and the link ${\rm s}\!\rightarrow\!{\rm sd}$ is not in outage. The mean service rate of $Q_{\rm s}$ is then given by
\begin{equation}
\begin{split}
  \mu_{\rm s} &\!=\!  \pi_\circ \Bigg[\!\omega\exp(-\!\frac{2^{\frac{b}{\delta W (T\!-\!\tau)}}\!-\!1}{\sigma_{{\rm s},{\rm sd}}\gamma_{{\rm s},{\rm sd}}})\!+\!\overline{\omega}\exp(-\!\frac{2^{\frac{b}{(1\!-\!\delta) W (T\!-\!\tau) }}\!-\!1}{\sigma_{{\rm s},{\rm sd}}\gamma_{{\rm s},{\rm sd}}})\!\Bigg]
    \end{split}
 \end{equation}
In a similar fashion, the mean service rate of $Q_{\rm ps}$ is given by
 \begin{equation}
\begin{split}
  \mu_{\rm ps} &\!=\! \pi_\circ\ \Bigg[\!\omega\exp(-\frac{2^{\frac{b}{(1\!-\!\delta) W (T\!-\!\tau)}}\!-\!1}{\sigma_{{\rm s},{\rm pd}}\gamma_{{\rm s},{\rm pd}}})\!+\!\overline{\omega}\exp(-\frac{2^{\frac{b}{\delta W (T\!-\!\tau) }}\!-1}{\sigma_{{\rm s},{\rm pd}}\gamma_{{\rm s},{\rm pd}}}\!)\!\Bigg]
  \end{split}
 \end{equation}

We present below two optimization problems to obtain $\omega$ and $\delta$. For strictly stationary arrival and service processes, a queue $Q$ with mean arrival rate $\lambda$ and mean service rate $\mu$ is stable if $\mu\ge \lambda$ \cite{6678828,sadek}. Once the SU obtains $\omega$, it determines probabilistically the subband allocation of the time slots. The generated schedule is then broadcasted to the primary and secondary receivers so that each knows which subband to decode at a particular time slot. We can operate without transmitting the schedule but with the cost of decoding both subbands. In this case, the receivers attempt to decode the transmission over the possible subbands, $W_1$ and $W_2$, and then select the correct decoding based on Cyclic Check Redundancy (CRC) appended to the packet.

\subsection{First Formulation: Throughput Maximization}
 The secondary mean service rate is maximized as $\delta$ and $\omega$ vary over $[0,1]$ and under the constraints of the stability of all other queues in the system. The optimization problem can be stated as follows:
 \begin{equation}
\begin{split}
\begin{split}
& \underset{\substack{0\le\omega,\delta\le1}}{\max.} \,\,\ \mu_{\rm s},\\ &
\,\,{\rm s.t.} \,\,\,\,\,\,\,\,\,\,\,\,\,\,\ \lambda_{\rm p} \le \mu_{\rm p}, \ \lambda_{\rm ps} \le \mu_{\rm ps}
\end{split}
\label{optoptbef}
\end{split}
\end{equation}

For a given $\delta$, the optimization problem is a linear program and can be readily solved as follows. We note that $\exp(-\frac{2^{\frac{b}{\delta W (T\!-\!\tau)}}\!-1}{\sigma_{{\rm s},{\rm sd}}\gamma_{{\rm s},{\rm sd}}})$ and $\exp(-\frac{2^{\frac{b}{\delta W (T\!-\!\tau) }}\!-1}{\sigma_{{\rm s},{\rm pd}}\gamma_{{\rm s},{\rm pd}}})$ are monotonic in $\delta$. Let $\zeta_n\!=\!\frac{\lambda_{\rm ps}}{\pi_\circ} \!-\! \exp(-\frac{2^{\frac{b}{\delta_n W (T\!-\!\tau) }}\!-1}{\sigma_{{\rm s},{\rm pd}}\gamma_{{\rm s},{\rm pd}}})$, $n\in\{1,2\}$, and $\beta \!=\!\zeta_1\!-\!\zeta_2$. For a given $\delta$ and $\lambda_{\rm p}\!\le\! \mu_{\rm p}$, the optimization problem can be rewritten as follows:

 \begin{equation}
\begin{split}
\begin{split}
& \underset{\substack{0\le\omega\le1}}{\max.} \,\,\ \eta \omega ,\\ &
\,\,{\rm s.t.} \,\,\,\,\,\,\,\,\ \zeta_1 \le \beta \omega
\end{split}
\label{optoptbef2}
\end{split}
\end{equation}
where  $\eta=\exp(-\frac{2^{\frac{b}{\delta W (T\!-\!\tau) }}\!-1}{\sigma_{{\rm s},{\rm sd}}\gamma_{{\rm s},{\rm sd}}})-\exp(-\frac{2^{\frac{b}{(1\!-\!\delta) W (T\!-\!\tau) }}\!-1}{\sigma_{{\rm s},{\rm sd}}\gamma_{{\rm s},{\rm sd}}})$. Note that if $\delta\!>\!1/2$, $\beta\!<\!0$ and $\eta\!>\!0$, otherwise $\beta\!>\!0$ and $\eta\!<\!0$. The optimal $\omega$, $\omega^*$, as a function of $\delta$ is given as follows:
\begin{itemize}
    \item If $\delta \!>\!1/2\!$ and $\zeta_1\!\le\! 0$, $\omega^*\!=\!\min(\frac{\zeta_1}{ \beta},1)$.
    \item If $\delta \!<\!1/2$ and $\zeta_2\!\le\!0$,  $\omega^*=\max(\frac{\zeta_1}{\beta},0)$.
        \item If $\delta \!=\!1/2$, the optimization problem becomes a feasibility problem.
        \item If $\delta\!<\! 1/2$ and $\zeta_1\!>\! 0$; or $\delta \!<\!1/2$ and $\zeta_2\!>\!0$, the problem is \textbf{infeasible}.
        \end{itemize}
with $\lambda_{\rm p}\!\le\! \mu_{\rm p}$. Note that $\min(\cdot,\cdot)$ and $\max(\cdot,\cdot)$ return the minimum and the maximum of the values in the argument, respectively. The first case can be explained as follows. Since $\delta>1/2$, $\beta$ is negative and $\eta$ is positive. Hence, maximizing $\eta\omega$ is equivalent to maximizing $\omega$ given the constraint that $\omega \!\leq\! \frac{\zeta_1}{\beta}$. Since $\zeta_1$ is also negative, then $\omega^*=\min(\frac{\zeta_1}{ \beta},1)$. The other cases can be explained in a similar fashion. We solve a family of linear programs parameterized by $\delta$. The optimal $\delta$ is obtained via a simple grid search over the set $[0,1]$ and is taken as the one which yields the highest objective function in (\ref{optoptbef}).

To provide further insights for the proposed protocol under this formulation, in the following subsections, we investigate two special cases.
\subsubsection{The case of $\omega\!=\!1$}

We investigate here the first special case of the proposed protocol, where $\omega$ is set to unity.\footnote{Equivalently, we can set $\omega$ to zero.} That is, the bandwidth assignment to queues is deterministic; $Q_{\rm s}$ is assigned $W_1\!=\!\delta W$ and $Q_{\rm ps}$ is assigned $W_2\!=\!(1-\delta)W$ of the total bandwidth. The service rate of the primary queue and the arrival rate of the relaying queue are given in (\ref{xor}) and (\ref{lambdaps}), respectively. The mean service rate of the secondary queues are given by
    \begin{equation}
\begin{split}
  \mu_{\rm s} &\!=\! \pi_\circ \exp(-\frac{2^{\frac{b}{(T\!-\!\tau)W\delta }}\!-\!1}{\gamma_{\rm s,sd}\sigma_{\rm s,sd}}) ,\\ \mu_{\rm ps} &= \pi_\circ \exp(-\frac{2^{\frac{b}{(T\!-\!\tau)W(1-\delta) }}\!-\!1}{\gamma_{\rm s,pd}\sigma_{\rm s,pd}})
  \label{case1}
  \end{split}
 \end{equation}

 Using $\mu_{\rm s}$ and $\mu_{\rm ps}$ in (\ref{case1}), $\mu_{\rm p}$ in (\ref{xor}), and $\lambda_{\rm ps}$ in (\ref{lambdaps}), the optimization problem in this case is given by
 \begin{equation}
\begin{split}
& \underset{\substack{\delta \in [0,1]}}{\max.} \,\,\ \!\mu_{{\rm s}}= \pi_\circ \exp(-\frac{2^{\frac{b}{(T\!-\!\tau)W\delta }}\!-\!1}{\gamma_{\rm s,sd}\sigma_{\rm s,sd}}),\\ &
\,\,{\rm s.t.} \,\,\,\,\ \lambda_{\rm p} \!\le\! \mu_{\rm p}, \ \lambda_{\rm ps} \!\le\! \mu_{\rm ps}
\end{split}
\label{opi2}
\end{equation}
The stability constraint of the relaying queue becomes
\begin{equation}
\begin{split}
\frac{\overline{\mathbb{P}_{{\rm p},{\rm s}}} \mathbb{P}_{\rm p,pd}\lambda_{\rm p}}{\mu_{\rm p}\!-\!\lambda_{\rm p}}\!\le  \!\exp(-\frac{2^{\frac{b}{(T\!-\!\tau)W(1-\delta) }}\!-\!1}{\gamma_{\rm s,pd}\sigma_{\rm s,pd}}) \!\le\! \exp(\!-\!\frac{2^{\frac{b}{(T\!-\!\tau)W}}\!-\!1}{\gamma_{\rm s,pd}\sigma_{\rm s,pd}})
  \end{split}
\end{equation}
The last inequality holds to equality when $\delta=0$, with
\begin{equation}
\frac{\overline{\mathbb{P}_{{\rm p},{\rm s}}} \mathbb{P}_{\rm p,pd}\lambda_{\rm p}}{\mu_{\rm p}-\lambda_{\rm p}}\!\le\! \exp(-\!\frac{2^{\frac{b}{(T\!-\!\tau)W}}\!-\!1}{\gamma_{\rm s,pd}\sigma_{\rm s,pd}})
\end{equation}
which implies that
\begin{equation}
\label{btex}
\lambda_{\rm p}\!\le\! \lambda^{\max}_{\rm p}\!=\!\frac{\mu_{\rm p}\exp(-\!\frac{2^{\frac{b}{(T\!-\!\tau)W}}\!-\!1}{\gamma_{\rm s,pd}\sigma_{\rm s,pd}})}{\exp(-\!\frac{2^{\frac{b}{(T\!-\!\tau)W}}\!-\!1}{\gamma_{\rm s,pd}\sigma_{\rm s,pd}})\!+\!\overline{\mathbb{P}_{{\rm p},{\rm s}}}\mathbb{P}_{\rm p,pd}}\!\le\!\mu_{\rm p}
\end{equation}
The stability of the primary queue is attained when $\lambda_{\rm p}~\le~\mu_{\rm p}$. This condition is tacitly included in constraint (\ref{btex}).
After some mathematical manipulations, the relaying queue stability constraint can be rewritten as
 \begin{equation}
\begin{split}
\label{xxds12}
(1\!-\!\delta) &\!\ge\! \frac{\mathcal{R}}{(1\!-\!\frac{\tau}{T})\log_2\Big[\!1\!+\!{\gamma_{\rm s,pd}\sigma_{\rm s,pd}}\ln\Big(\!\frac{\mu_{\rm p}-{\lambda_{\rm p}}}{\overline{\mathbb{P}_{{\rm p},{\rm s}}}\mathbb{P}_{\rm p,pd}\lambda_{\rm p}}\!\Big)\!\Big]\!}\!=\! \kappa
  \end{split}
\end{equation}
with $0\!\le\! \lambda_{\rm p}\!\le\! \lambda^{\max}_{\rm p}$ and $\kappa\le1$. If $\kappa>1$, and since $\delta \in [0,1]$, the optimization problem (\ref{opi2}) is infeasible, i.e., the relaying queue cannot be maintained stable for any feasible value of $\delta$.

The optimization problem (\ref{opi2}) can be converted to a convex program by taking the logarithm of the objective function and the constraint. After some simplifications, the optimization problem is given by
 \begin{equation}
\begin{split}
 \underset{\substack{\delta \in [0,1]}}{\min.}  & \  2^{\frac{b}{(T\!-\!\tau)W\delta }},\
{\rm s.t.} \  (1-\delta)\ge \kappa
\end{split}
\end{equation}
If $\kappa\le 1$, and since the objective function is monotonically decreasing with $\delta$, the minimum of the objective function is attained when $1-\delta$ is adjusted to its lowest feasible value, $\kappa$. Hence, the optimal fraction of the bandwidth assigned to the relaying queue which achieves the maximum secondary throughput and maintains the stability of both the primary and the relaying queues is given by
 \begin{equation}
\begin{split}
\label{optee}
1\!-\!\delta^* \!=\!\kappa\!=\! \frac{\mathcal{R}}{(1\!-\!\frac{\tau}{T})\log_2\Biggr[\!1\!+\!{\gamma_{\rm s,pd}\sigma_{\rm s,pd}}\ln\Big(\!\frac{\mu_{\rm p}-{\lambda_{\rm p}}}{\overline{\mathbb{P}_{{\rm p},{\rm s}}}\mathbb{P}_{\rm p,pd}\lambda_{\rm p}}\!\Big)\!\Biggr]}
  \end{split}
\end{equation}
with $0\!\le\! \lambda_{\rm p}\!\le\! \lambda^{\max}_{\rm p}$ and $\kappa\le1$. As ${\gamma_{\rm s,pd}\sigma_{\rm s,pd}}$ increases, the bandwidth assigned to the relaying packets, $(1-\delta^*)  W$, decreases. This is because if the SNR is high, the probability of successful packet decoding at the destination is high as well. Hence, it is better in terms of the secondary throughput to assign more bandwidth to the secondary packet for increasing its successful decoding probability (throughput). Moreover, as the primary mean arrival, or the rate $\mathcal{R}$ increases, $(1-\delta^*) $ increases as well. This has the following intuitive explanation: As the mean arrival rate or the primary channel outage increases, the number of primary packets flowing to the relaying queue increases as well. Hence, to maintain the stability of the relaying queue, the probability of correct packet reception should be increased to boost its mean service rate. Furthermore, as $\mathcal{R}$ increases, the outage probability of the link between the relaying queue and the primary destination increases. Hence, the bandwidth assigned to the relaying queue should be increased in order to decrease the probability of channel outage and maintain the primary and relaying queues stability.

Using (\ref{optee}), the maximum spectral efficiency rate of the primary system can be obtained by setting $\delta$ to zero (or assigning all bandwidth to the relaying queue all the time). Specifically, the maximum spectral rate, $\mathcal{R}_{\max}$, that can be used by the primary system, when the SU is available to assist, is achieved for a given $0\!\le\! \lambda_{\rm p}\!\le\! \lambda^{\max}_{\rm p}$ packets of size $b$ per time slot when $\delta\!=\!0$. That is,
 \begin{equation}
\begin{split}
\frac{\mathcal{R}_{\max}}{(1\!-\!\frac{\tau}{T})\log_2\Biggr[\!1\!+\!{\gamma_{\rm s,pd}\sigma_{\rm s,pd}}\ln\Big(\!\frac{\mu_{\rm p}-{\lambda_{\rm p}}}{\overline{\mathbb{P}_{{\rm p},{\rm s}}}\mathbb{P}_{\rm p,pd}\lambda_{\rm p}}\!\Big)\!\Biggr]}\!=\! 1
  \end{split}
\end{equation}
Thus,
 \begin{equation}
\begin{split}
{\mathcal{R}_{\max}}\!=\!{(1\!-\!\frac{\tau}{T})\log_2\Biggr[\!1\!+\!{\gamma_{\rm s,pd}\sigma_{\rm s,pd}}\ln\Big(\!\frac{\mu_{\rm p}-{\lambda_{\rm p}}}{\overline{\mathbb{P}_{{\rm p},{\rm s}}}\mathbb{P}_{\rm p,pd}\lambda_{\rm p}}\!\Big)\!\Biggr]}
  \end{split}
\end{equation}
with $0\!\le\! \lambda_{\rm p}\!\le\! \lambda^{\max}_{\rm p}$. The achievable rate increases with the received SNR and decreases with $\lambda_{\rm p}$ and $\tau/T$. The impact of $\mu_{\rm p}$ and $\overline{\mathbb{P}_{{\rm p},{\rm s}}}\mathbb{P}_{\rm p,pd}$ on ${\mathcal{R}_{\max}}$ cannot be determined because it depends on the relationship between those parameters and the others in the system.

\subsubsection{PCR protocol}
In this subsection, we investigate the maximum stable throughput of the PCR, which has been studied in several works such as \cite{khattab} and \cite{rong2012cooperative}, and prove that its stable throughput can be achieved by the proposed protocol. In PCR system, when the PU is inactive, the SU retransmits a packet from the relaying queue, $Q_{\rm ps}$, with transmission bandwidth $W$. When both the primary and the relaying queues are empty, the SU transmits a packet from its own queue, $Q_{\rm s}$, with transmission bandwidth $W$. When the PU is active, the SU remains silent and attempts to decode the primary packet and store it if the primary destination fails to decode it. The probability of the relaying queue being empty is given by

 \begin{equation}
\begin{split}
{\rm Pr}\{Q_{\rm ps}=0\}\!=\!1\!-\!\frac{\lambda_{\rm ps}}{\mu_{\rm ps}}\!=\!1\!-\!\frac{\overline{\mathbb{P}_{{\rm p},{\rm s}}}\mathbb{P}_{\rm p,pd}\overline{\pi_\circ}}{ \pi_\circ \exp(\!-\frac{2^{\frac{b}{W (T\!-\!\tau)}}\!-\!1}{\sigma_{{\rm s},{\rm pd}}\gamma_{{\rm s},{\rm pd}}})}
\end{split}
 \end{equation}
 The maximum secondary stable throughput is given by
 \begin{equation}
\begin{split}
 \lambda_{\rm s}\!\le\!\mu_{\rm s}\!=\! (\!1\!-\!\frac{\overline{\mathbb{P}_{{\rm p},{\rm s}}}\mathbb{P}_{\rm p,pd}\overline{\pi_\circ}}{ \pi_\circ \exp(\!-\frac{2^{\frac{b}{W (T\!-\!\tau)}}\!-\!1}{\sigma_{{\rm s},{\rm pd}}\gamma_{{\rm s},{\rm pd}}})})\pi_\circ \ \exp(\!-\frac{2^{\frac{b}{W (T\!-\!\tau)}}\!-1}{\sigma_{{\rm s},{\rm sd}}\gamma_{{\rm s},{\rm sd}}}\!)
 \label{ddfs}
  \end{split}
 \end{equation}
with $0\le \lambda_{\rm p}\le \lambda^{\max}_{\rm p}$.

 The maximum secondary stable throughput under the PCR protocol is achieved under the proposed protocol in this paper when
\begin{equation}
\begin{split}
\delta\!=\!1, \ \omega\!=\! 1\!-\! \frac{\overline{\mathbb{P}_{{\rm p},{\rm s}}}\mathbb{P}_{\rm p,pd}\overline{\pi_\circ}}{ \pi_\circ \exp(-\frac{2^{\frac{b}{W (T\!-\!\tau)}}\!-1}{\sigma_{{\rm s},{\rm pd}}\gamma_{{\rm s},{\rm pd}}})}
\end{split}
\end{equation}
with $0\!\le\! \lambda_{\rm p}\!\le\! \lambda^{\max}_{\rm p}$. Since the maximum secondary stable throughput under the PCR protocol is an achievable throughput under the proposed protocol, the proposed protocol outperforms the PCR protocol.

\subsection{Second Formulation: Minimum Secondary Queueing Delay}
Let $\mu_{\rm s}=\pi_\circ \phi_{\rm s,sd}$ and $\mu_{\rm ps}=\pi_\circ \phi_{\rm s,pd}$, where \begin{equation}
\begin{split}
\label{phis}
  \phi_{\rm s,sd} &\!=\! \omega\exp(-\frac{2^{\frac{b}{\delta W (T\!-\!\tau)}}\!-1}{\sigma_{{\rm s},{\rm sd}}\gamma_{{\rm s},{\rm sd}}})\!+\!\overline{\omega}\exp(-\frac{2^{\frac{b}{(1\!-\!\delta) W (T\!-\!\tau) }}\!-1}{\sigma_{{\rm s},{\rm sd}}\gamma_{{\rm s},{\rm sd}}})\\
  \phi_{\rm s,pd} &\!=\!\omega\exp(-\frac{2^{\frac{b}{(1\!-\!\delta) W (T\!-\!\tau)}}\!-\!1}{\sigma_{{\rm s},{\rm pd}}\gamma_{{\rm s},{\rm pd}}})\!+\!\overline{\omega}\exp(-\frac{2^{\frac{b}{\delta W (T\!-\!\tau) }}\!-1}{\sigma_{{\rm s},{\rm pd}}\gamma_{{\rm s},{\rm pd}}})
    \end{split}
 \end{equation}
 Queuing delays can be obtained using the same moment generating function approach employed in \cite{rong2012cooperative}. The secondary queueing delay, $D_{\rm s}$, follows \cite[Eqn. 23]{rong2012cooperative} and is given by
     \begin{equation}
\begin{split}
D_{{\rm s}}=\frac{(-\mu_{\rm p}+\phi_{\rm s,sd}-\mu_{\rm p}\phi_{\rm s,sd})\lambda_{\rm p}-\mu_{\rm p}^2\lambda_{\rm s}+\mu_{\rm p}\lambda_{\rm p} \lambda_{\rm s}+\mu_{\rm p}^2}{(\phi_{\rm s,sd}\lambda_{\rm p}+\mu_{\rm p}\lambda_{\rm s}-\mu_{\rm p}\phi_{\rm s,sd})(\lambda_{\rm p}-\mu_{\rm p})}
    \end{split}
\end{equation}
The primary end-to-end queueing delay, $D_{\rm p}$, follows \cite[Eqn. 13]{rong2012cooperative}. That is, when the system is stable, the primary queueing delay, $D_{\rm p}$, is given by
     \begin{equation}
\begin{split}
D_{\rm p}=\frac{1-\lambda_{\rm p}}{\mu_{\rm p}-\lambda_{\rm p}}+\frac{f\lambda_{\rm p}+g}{a\lambda_{\rm p}^2+B\lambda_{\rm p}+c}
\label{Dp}
    \end{split}
\end{equation}
with $\mathcal{Y}=\mu_{\rm p}\!-\!\overline{\mathbb{P}_{\rm p,pd}}$,
    \begin{equation}
\begin{split}
f&\!=\!\mathcal{Y}(\frac{\phi_{\rm s,pd}\!-\!\overline{\mathbb{P}_{\rm p,pd}}}{\mu_{\rm p}}-a), g\!=\!\mathcal{Y}\mu_{\rm p},\\
a&=\mathcal{Y}+\phi_{\rm s,pd},\ B=\mu_{\rm p}(-a-\phi_{\rm s,pd}), c=\phi_{\rm s,pd}\mu_{\rm p}^2
 \end{split}
\end{equation}
 The first term in (\ref{Dp}), ${(1-\lambda_{\rm p})}/{(\mu_{\rm p}-\lambda_{\rm p})}$, represents the delay that a packet stored at $Q_{\rm p}$ would suffer from, while the second term in (\ref{Dp}) represents the average queue length of $Q_{\rm ps}$ normalized by $\lambda_{\rm p}$.

 The minimum secondary queueing delay for a given arrival rates pair $(\lambda_{\rm p},\lambda_{\rm s})$, if the system is stable and under certain tolerable primary queueing delay, $D_{\rm p}\le \mathcal{D}$, is obtained via solving the following optimization problem:
 \begin{equation}
\begin{split}
\begin{split}
& \underset{\substack{0\le\omega,\delta\le1}}{\min.} \,\,\ D_{\rm s},\\&
\,\,{\rm s.t.} \,\,\,\,\,\,\,\,\,\,\,\,\ \lambda_{\rm p} \!\le\! \mu_{\rm p}, \lambda_{\rm s} \!\le\! \mu_{\rm s}, \ \lambda_{\rm ps} \!\le\! \mu_{\rm ps},
\\& \,\,\,\,\,\,\,\,\,\,\,\,\,\,\,\,\,\,\,\,\,\,\,\ D_{\rm p}\!\le\! \mathcal{D}
\label{opt2d}
\end{split}
\end{split}
\end{equation}
The constraints $\lambda_{\rm p} \le \mu_{\rm p}$, $\lambda_{\rm s} \le \mu_{\rm s}$ and $\lambda_{\rm ps} \le \mu_{\rm ps}$ represent the system stability and the constraint $D_{\rm p}\le \mathcal{D}$, where $\mathcal{D}$ is a specific application-based delay constraint, represents certain QoS requirement for the PU.
The primary end-to-end queueing delay constraint, $D_{\rm p}\le\mathcal{D}$, can be rewritten as
     \begin{equation}
\begin{split}
\label{cko}
\frac{\mathcal{Y}\lambda_{\rm p} \phi_{\rm s,pd}(\frac{1}{\mu_{\rm p}}\!-\! 1)\!-\!\mathcal{Y}\lambda_{\rm p} (\frac{\overline{\mathbb{P}_{\rm p,pd}}}{\mu_{\rm p}}+\mathcal{Y})+g}{(\mu_{\rm p}-\lambda_{\rm p})\mathbb{D}[\phi_{\rm s,pd}(\mu_{\rm p}-\lambda_{\rm p})-\mathcal{Y}\lambda_{\rm p}]}\le1
    \end{split}
\end{equation}
where $\mathbb{D}=\mathcal{D}-{(1-\lambda_{\rm p})}/({\mu_{\rm p}-\lambda_{\rm p}})$. The numerator of (\ref{cko}) is always positive if $g=\mathcal{Y}\mu_{\rm p}\ge\!\mathcal{Y}\lambda_{\rm p} (\frac{\overline{\mathbb{P}_{\rm p,pd}}}{\mu_{\rm p}}+\mathcal{Y})$. This condition is always satisfied as far as $Q_{\rm p}$ is stable, i.e., $ \lambda_{\rm p}\le \mu_{\rm p}$. Moreover, the denominator of (\ref{cko}) is positive for $\lambda_{\rm ps}\le \mu_{\rm ps}$. This condition is always satisfied as far as the relaying queue is stable, i.e., $\lambda_{\rm ps}\le \mu_{\rm ps}$. (\ref{cko}) is simplified to
     \begin{equation}
\begin{split}
{\phi_{\rm s,pd}\Psi}\le\mathcal{J}, \mathcal{J}=-{\mathcal{Y}\mathbb{D}\lambda_{\rm p}(\mu_{\rm p}-\lambda_{\rm p})\!+\!\mathcal{Y}\lambda_{\rm p} (\frac{\overline{\mathbb{P}_{\rm p,pd}}}{\mu_{\rm p}}+\mathcal{Y})-g}
    \end{split}
\end{equation}
where $\Psi\!=\!\mathcal{Y}\lambda_{\rm p} (\frac{1}{\mu_{\rm p}}\!-\! 1)\!-\!\mathbb{D}(\mu_{\rm p}\!-\!\lambda_{\rm p})^2$.
For a given $\delta$ and $ \lambda_{\rm p} \!\le\! \mu_{\rm p}$, the optimization problem (\ref{opt2d}) can be rewritten as
 \begin{equation}
\begin{split}
\begin{split}
\label{gokooo}
&\underset{\substack{0\le\omega\le1}}{\min.} \,\,\ \frac{\overline{\mu_{\rm p}}\lambda_{\rm p}\omega\eta+\overline{\mu_{\rm p}}\lambda_{\rm p}\exp(-\frac{2^{\frac{b}{(1\!-\!\delta) W (T\!-\!\tau) }}\!-1}{\sigma_{{\rm s},{\rm sd}}\gamma_{{\rm s},{\rm sd}}})\!+\!\overline{\lambda_{\rm s}}\mu^2_{\rm p}\pi_\circ}{\omega\eta\pi_\circ+\pi_\circ\exp(-\frac{2^{\frac{b}{(1\!-\!\delta) W (T\!-\!\tau) }}\!-1}{\sigma_{{\rm s},{\rm sd}}\gamma_{{\rm s},{\rm sd}}})-\lambda_{\rm s}}\! \\ &
\,\,{\rm s.t.} \ \frac{\lambda_{\rm s}}{\pi_\circ}\!-\!\exp(-\frac{2^{\frac{b}{(1\!-\!\delta) W (T\!-\!\tau) }}\!-1}{\sigma_{{\rm s},{\rm sd}}\gamma_{{\rm s},{\rm sd}}}) \!\le\! \eta \omega, \ \zeta_1 \le \beta \omega,\\ & \,\,\,\,\,\,\,\,\ \omega\beta \Psi\le\mathcal{J}-\Psi \exp(-\frac{2^{\frac{b}{\delta W (T\!-\!\tau) }}\!-1}{\sigma_{{\rm s},{\rm pd}}\gamma_{{\rm s},{\rm pd}}})
\end{split}
\end{split}
\end{equation}
The objective function (\ref{gokooo}) is linear-fractional on $\omega$. Since the objective function is linear-fractional and the constraints are linear, the optimization problem is a linear-fractional program. Linear fractional programs can be converted to linear programs as explained in \cite[page 151]{boyed}. We then solve a family of linear programs parameterized by $\delta$, which is obtained via a simple grid search over the set $[0,1]$. The optimal $\delta$ is chosen as the one which yields the lowest objective function in (\ref{opt2d}).
\section{Numerical Results and Simulations}\label{sec3}

\begin{figure}
  \centering
  \includegraphics[width=1\columnwidth]{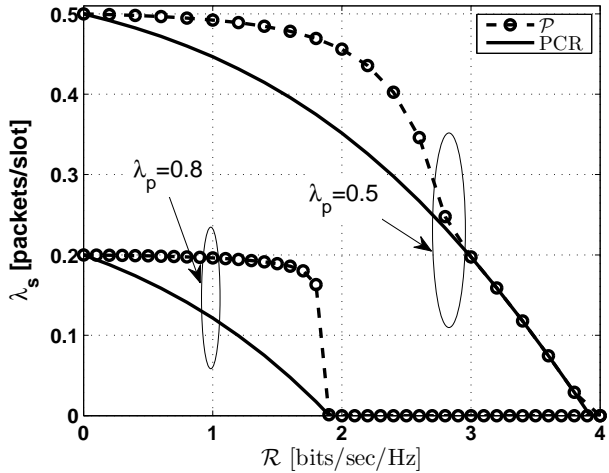}\\
   \caption{The maximum secondary stable throughput versus $\mathcal{R}$ for two values of $\lambda_{\rm p}$. The figure is generated using the following parameters: $\tau\!=\!0.1T$, $\sigma_{\rm p,s}\!=\!\sigma_{\rm p,pd}\!=\!\sigma_{\rm s,sd}\!=\!\sigma_{\rm s,pd}\!=\!1$, $P^{\rm s}\!=\!10^{-9}$ Watts/Hz, $P^{\rm p}\!=\!10^{-10}$ Watts/Hz, and $\mathcal{N}_\circ=10^{-11}$ Watts/Hz.
}\label{fig1}
\end{figure}
\begin{figure}
  \centering
  \includegraphics[width=1\columnwidth]{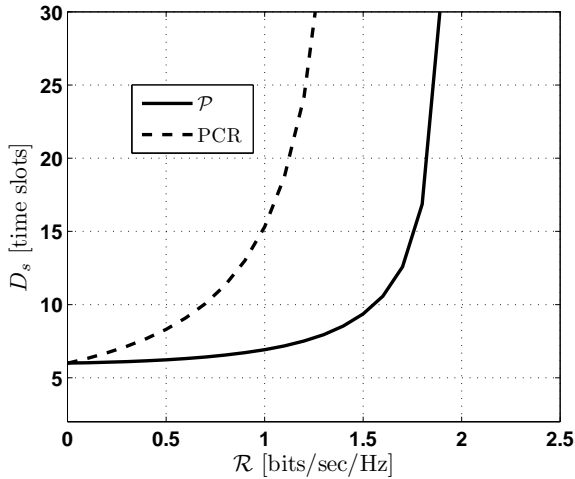}\\
   \caption{The minimum secondary queueing delay for $\tau\!=\!0.1T$, $\sigma_{\rm p,s}\!=\!\sigma_{\rm p,pd}\!=\!\sigma_{\rm s,sd}\!=\!\sigma_{\rm s,pd}\!=\!1$, $P^{\rm s}\!=\!10^{-9}$ Watts/Hz, $P^{\rm p}\!=\!10^{-10}$ Watts/Hz, $\mathcal{N}_\circ=10^{-11}$ Watts/Hz, $\lambda_{\rm p}=0.5$ packets per time slot and $\lambda_{\rm s}=0.4$ packets per time slots. The figure is generated with maximum primary end-to-end queueing delay $\mathcal{D}\!=\!2$ time slots. That is, $D_{\rm p}\le2$ time slots.
}\label{fig2}
\end{figure}
Some numerical results are presented in this section. Let $\mathcal{P}$ denote the proposed cooperative cognitive protocol. The figures are generated using the following common parameters: $\tau\!=\!0.1T$, $\sigma_{\rm p,s}\!=\!\sigma_{\rm p,pd}\!=\!\sigma_{\rm s,sd}\!=\!\sigma_{\rm s,pd}\!=\!1$, $P^{\rm s}\!=\!10^{-9}$ Watts/Hz, $P^{\rm p}\!=\!10^{-10}$ Watts/Hz, and $\mathcal{N}_\circ=10^{-11}$ Watts/Hz.
 Fig. \ref{fig1} shows the maximum secondary stable throughput versus rate $\mathcal{R}$. The figure shows the monotonicity of $\lambda_{\rm s}$ with $\mathcal{R}$ and $\lambda_{\rm p}$. It is noted in the figure that for $\lambda_{\rm p}\!=\!0.5$ packets/slot in low $\mathcal{R}$ regimes, $\mathcal{P}$ outperforms PCR. While in high $\mathcal{R}$ regimes, $\mathcal{P}$ coincides with PCR. The figure also shows that at high $\lambda_{\rm p}$, $\mathcal{P}$ always outperforms PCR. Specifically, for $\lambda_{\rm p}\!=\!0.8$ packets/slot, $\mathcal{P}$ outperforms PCR for all $\mathcal{R}$.
Fig. \ref{fig2} shows the advantage of cooperation over the non-cooperation case in terms of the secondary queueing delay. As shown in the plot, the proposed protocol provides better queueing delay over the PCR protocol.
The figure is generated with a primary queueing delay $D_{\rm p}\le 2$ time slots, $\lambda_{\rm p}=0.5$ packets per time slot and $\lambda_{\rm s}=0.4$ packets per time slot.

\section{Conclusions}\label{sec4}
In this paper, we have proposed a new cooperative protocol which involves cooperation between the PUs and the SUs. The SU probabilistically splits the bandwidth assigned to its own queue and the relaying queue. We have compared the proposed system with the PCR system, and showed that the proposed system outperforms the PCR system. We have derived the stability region of the proposed protocol. We have also derived the end-to-end queueing delay expressions for the PU and the SU. Moreover, we have proposed a formulation which minimizes the secondary queueing delay subject to stability constraints of the queues and certain quality of service requirement on the primary end-to-end queueing delay.

We are currently investigating a system in which the PU releases a portion of its bandwidth and time slot duration for the SU. In this case, the SU uses the channel continuously, and it simultaneously transmits its packets with the PU each time slot. The SU probabilistically splits the released bandwidth among the relaying queue and its own data queue. We also include the impact of having the transmit CSI at the secondary transmitter to make the bandwidth assignment among secondary queue dynamic from slot to slot and depends on the instantaneous estimated channels gain. The imperfect estimation of the CSI of the secondary link and the link between the SU and the primary receiver at the secondary transmitter can also be taken into consideration.

\bibliographystyle{IEEEtran}
\bibliography{IEEEabrv,bibfile}
\balance
\end{document}